\documentclass[iop]{emulateapj}

\usepackage[sumlimits,intlimits]{amsmath}

\usepackage[english]{babel}

\shorttitle{Solar Modulation of Proton LIS}
\shortauthors{Corti et al.}

\begin{document}

\title{Solar Modulation of the Proton Local Interstellar Spectrum with AMS-02, Voyager 1 and PAMELA}

\author{C. Corti, V. Bindi, C. Consolandi and K. Whitman}
\affil{Physics and Astronomy Department, University of Hawaii at Manoa,
    Honolulu, HI 96822}
\email{corti@hawaii.edu}

\begin{abstract}
In recent years, the increasing precision of direct cosmic rays measurements opened the door to indirect searches of dark matter with high-sensitivity and to more accurate predictions for radiation doses received by astronauts and electronics in space. The key ingredients in the study of these phenomena are the knowledge of the local interstellar spectrum (LIS) of galactic cosmic rays (GCRs) and the understanding of how the solar modulation affects the LIS inside the heliosphere.
Voyager 1, AMS-02 and PAMELA measurements of proton fluxes provide invaluable information, allowing us to shed light on the shape of the LIS and the details of the solar modulation during solar cycles 23 and 24.
A new parametrization of the proton LIS is presented, based on the latest data from Voyager 1 and AMS-02. Using the framework of the force-field approximation, the solar modulation parameter is extracted from the time-dependent proton fluxes measured by PAMELA. A modified version of the force-field approximation with an energy-dependent modulation parameter is introduced, yielding better results on proton data than the force-field approximation. The results are compared with the modulation parameter inferred by neutron monitors.
\end{abstract}


\section{Introduction}
The search for the local interstellar spectrum (LIS) of galactic cosmic rays (GCRs) and a full understanding of the solar modulation are long-standing issues in the field of cosmic rays and heliophysics. In recent years, hints of possible dark matter (DM) signatures or new astrophysical phenomena have accumulated accumulated as a results of accurate measurements of the anti-matter component in cosmic ray fluxes \citep{bib:pos-pam,bib:posfrac-ams02,bib:pos-ams02,bib:pbar-pam}.
In order to define the astrophysical background of GCRs over which to look for the excess coming from DM annihilation or decay, the knowledge of the LIS is of utmost importance. Uncertainties in the low energy part of the LIS due to the solar modulation reduce the sensitivity of these type of searches \citep{bib:antiproton-solmod,bib:antideuteron-solmod,bib:positron-solmod}.
With the ever-growing number of satellites orbiting Earth and NASA plans for human missions to Mars, the characterization of the radiation dose received by astronauts and electronics in different periods of the solar cycle is becoming more and more important: a precise knowledge of the LIS and the temporal variation of GCR fluxes inside the heliosphere is needed for reducing the uncertainties on the estimated dose \citep{bib:dose-solmod,bib:oneill}.

Data collected over many decades from ground observations, balloon experiments and spacecraft have deepened our understanding of how the heliosphere affects the spectrum of GCRs: many numerical models have been developed to solve the Parker equation governing the propagation of GCRs in the heliosphere \citep{bib:parker} and to explore the different processes induced by their interactions between the heliospheric magnetic field and the solar wind. Nevertheless, the force-field approximation \citep{bib:ffa} is still routinely used as a reference, due to its simplicity. Under the assumptions of spherical symmetry, radial solar wind, an isotropic diffusion coefficient and no particle drift, the differential GCR flux $dJ/dT$, measured at Earth at the time $t$, is related to the LIS $dJ_{LIS}/dT$ via the formula

\begin{equation} \label{eqn:ffa}
   \frac{dJ}{dT}(T) = \frac{T(T + 2M)}{(T + \Phi)(T + \Phi + 2M)} \frac{dJ_{LIS}}{dT}(T + \Phi)
\end{equation}

where $T$ is the kinetic energy of a nucleus of charge $Z$ and mass $M$ and $\Phi = Z e \phi(t)$. $\phi(t)$ is known as the solar modulation parameter or solar modulation potential and has the units of an electric potential.

In August 2012, the Voyager 1 spacecraft, launched in 1977, crossed the heliopause and entered interstellar space \citep{bib:voyager-science}. A debate is still ongoing whether the heliopause can be considered the modulation boundary or not \citep{bib:hp-mod,bib:hp-maybe-mod,bib:hp-no-mod}, but so far the GCR flux measured by Voyager 1 has remained steady \footnote{See, for example, the proton rates from 2013 to 2015 at \url{http://voyager.gsfc.nasa.gov/heliopause/yearplot24h.html}}, thereby suggesting that what is being observed is actually a LIS.
In 2006, just before the minimum of solar cycle 23, the PAMELA experiment was launched on board a satellite in low Earth orbit and since has provided a precise and direct measurement of the top-of-atmosphere proton flux and its time variation up to 50 GeV \citep{bib:pamela-monthly-flux}. The AMS-02 experiment was installed in 2011 on the International Space Station during the ascending phase of solar cycle 24 and recently published the proton flux up to 2 TeV, integrated over 3 years, with an error at the $\%$ level \citep{bib:ams02-proton-flux}, which provides the most accurate measurement of the high energy part of the proton LIS. AMS-02 is expected to take data until the decommissioning of the ISS in 2024, allowing a precise measurement of the time variation of GCRs throughout an entire solar cycle and of the solar modulation effects on different species of cosmic rays.

In this paper, we provide a new parametrization for the proton LIS based on Voyager 1 and AMS-02 proton data. This new LIS model, modulated with the force-field approximation, is used to fit the monthly proton fluxes measured by PAMELA. We propose a modified version of the force-field approximation with an energy-dependent $\phi$ to better describe PAMELA data and finally, we compare the extracted $\phi(t)$ with the one derived from neutron monitors (NMs).

\section{A new parametrization for the proton LIS} \label{sec:new-lis}
\begin{figure}[!h]
\centering
\includegraphics[width=0.45\textwidth]{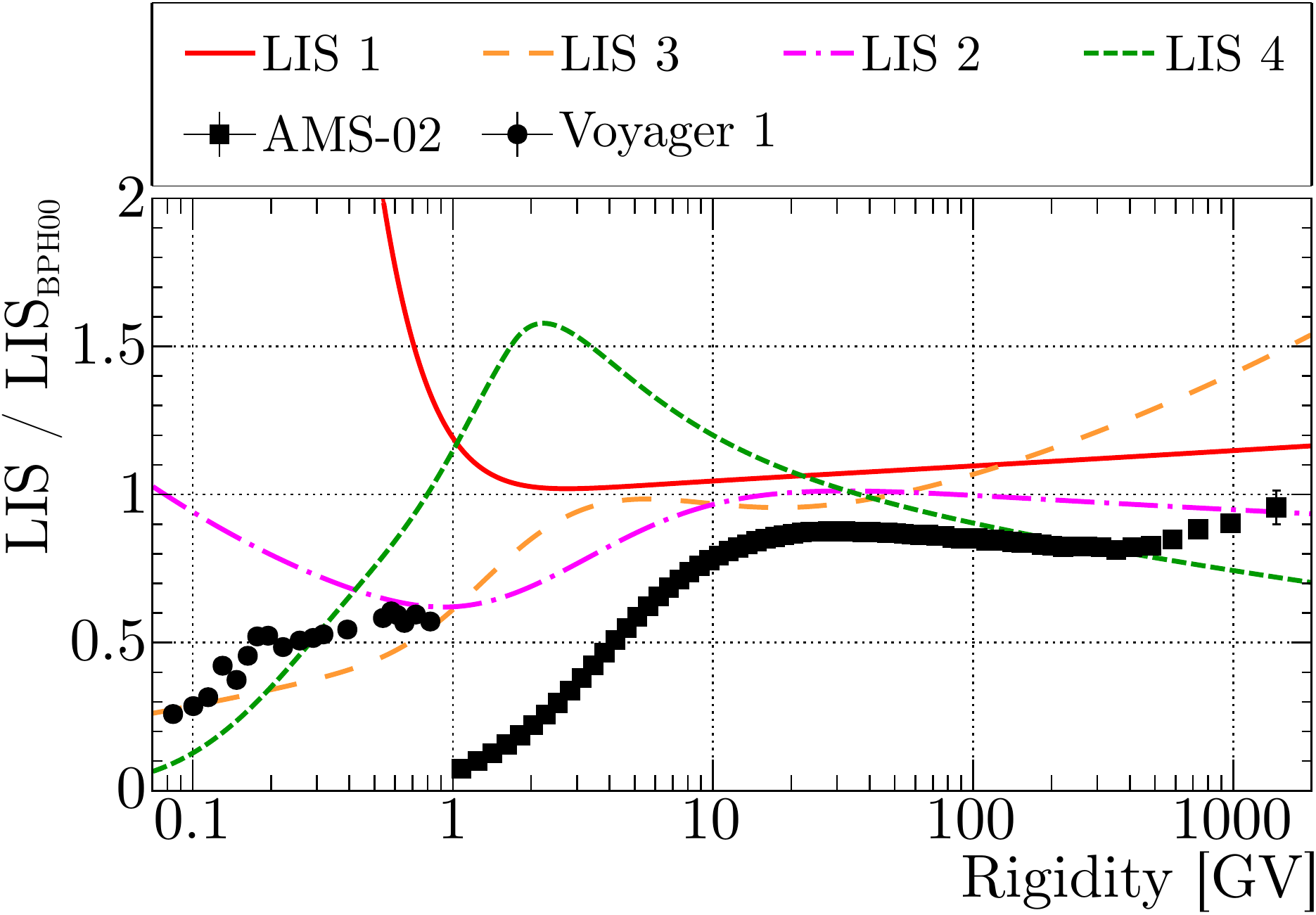}
\caption{Ratio of various proton LIS models (lines) and Voyager 1 (black dots) and AMS-02 proton (black squares) fluxes to the BPH00 model. The used models are: 1 \citep{bib:bess}; 2 \citep{bib:garcia-munoz}; 3 \cite{bib:langner}; 4 \citep{bib:webber-higbie}. A color version of this figure is available in the online journal.}
\label{fig:exp-lis-ratio}
\end{figure}

The majority of the LIS models found in literature are based on spacecraft and balloon measurements of GCRs before Voyager 1 entered the interstellar space and do not take into account a change of spectral index at high rigidities ($R \gtrsim 300$ GV), which has been observed by PAMELA \citep{bib:pamela-proton-flux} and AMS-02 \citep{bib:ams02-proton-flux}. The availability of the high-accuracy high energy proton flux from AMS-02 and the low energy proton flux from Voyager 1 represent important progress towards the reduction of the uncertainty on the LIS shape, enabling a more accurate determination of the solar modulation parameter and improving the understanding of GCR propagation in the heliosphere.
Figure \ref{fig:exp-lis-ratio} shows the ratio of various proton LIS models to the BPH00 model used in \citep{bib:Usoskin05} to extract the solar modulation parameter from NMs, along with the ratio of Voyager 1 \citep{bib:voyager-science} and AMS-02 \citep{bib:ams02-proton-flux} proton fluxes to the same model. It is clear that the new data from Voyager 1 and AMS-02 are not well described by these models. These discrepancies compel us to find a new LIS parametrization based on the new results from Voyager 1 and AMS-02.

For the high-energy end of the proton LIS, we use the model adopted by the AMS-02 collaboration \citep{bib:ams02-proton-flux} to describe a double power-law:

\begin{equation} \label{eqn:power-law-break}
   R^{\gamma} \left[ 1 + \left( \frac{R}{R_{b}} \right)^{\Delta\gamma/s} \right]^{s}
\end{equation}
where $\Delta\gamma$ is the change in spectral index, $R_{b}$ is the rigidity where the two power-laws cross each other and $s$ determines the smoothness of the change ($s=0$ means a broken power-law).
Recently, the analysis of $\gamma$-ray emissions from giant molecular clouds point to a low-energy break around 9 GeV, with the spectral index changing from $\approx\!-2$ to $\approx\!-3$ \citep{bib:neronov12}. We generalize equation \ref{eqn:power-law-break} to describe two power-law breaks:

\begin{equation} \label{eqn:two-power-law-break}
   R^{\gamma_{1}} \left\{ 1 + \left[ \frac{R}{R_{b1}} \left( 1 + \left( \frac{R}{R_{b2}} \right)^{\Delta\gamma_{2}/s_{2}} \right)^{s_{2}}\right]^{\Delta\gamma_{1}/s_{1}} \right\}^{s_{1}}
\end{equation}

where the indices 1 and 2 stand for the low- and high-rigidity break respectively: if $R \ll R_{b1}$, equation \ref{eqn:two-power-law-break} reduces to $\approx\!R^{\gamma_{1}}$; if $R_{b1} \ll R \ll R_{b2}$, equation \ref{eqn:two-power-law-break} becomes $\approx\!R^{\gamma_{1} + \Delta\gamma_{1}} = R^{\gamma_{2}}$; and if $R \gg R_{b2}$, equation \ref{eqn:two-power-law-break} goes as $\approx\!R^{\gamma_{1} + (1+\Delta\gamma_{2})\Delta\gamma_{1}} = R^{\gamma_{3}}$.

\begin{figure}[!h]
\centering
\includegraphics[width=0.45\textwidth]{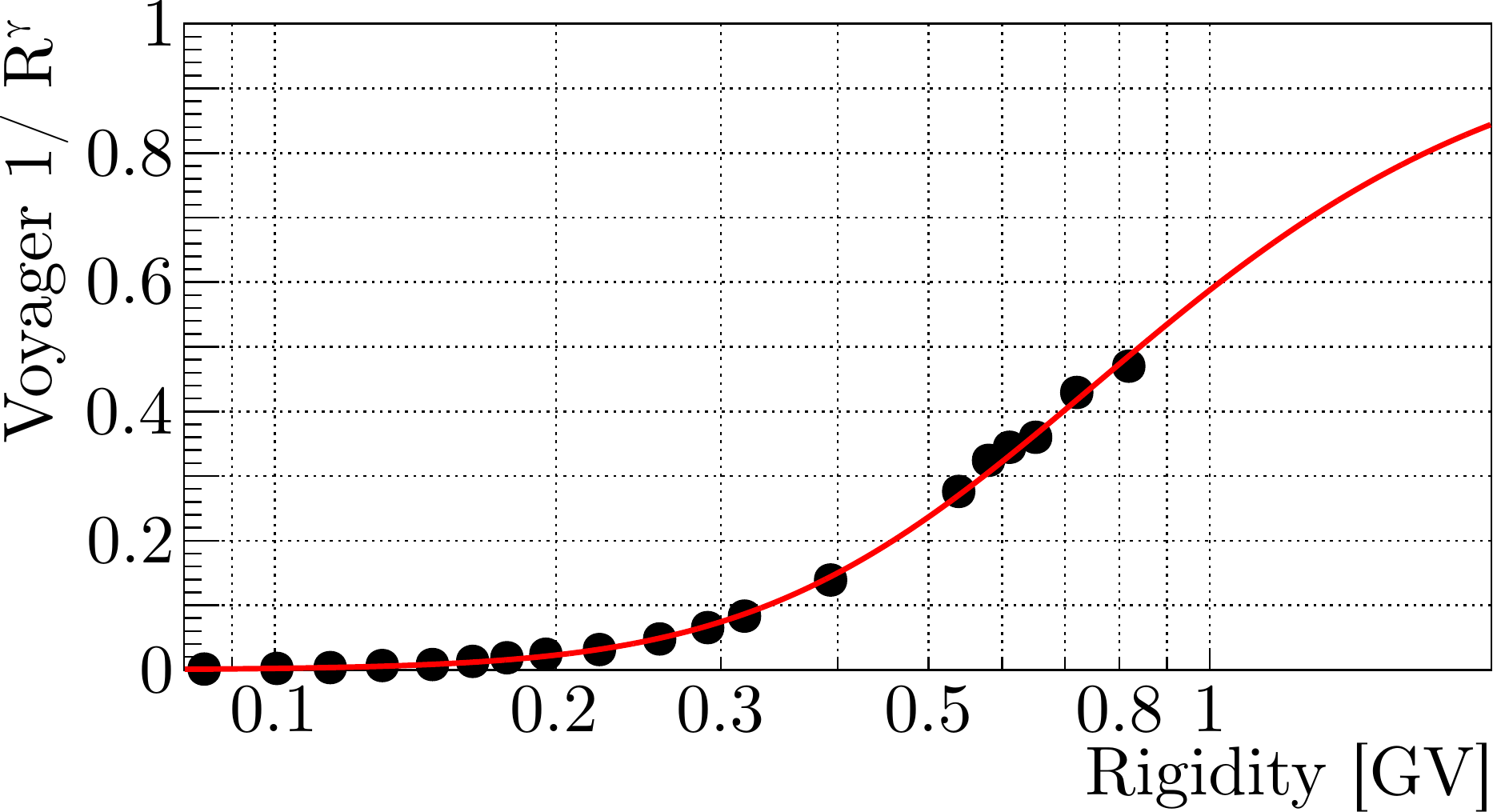}
\caption{Ratio of the proton flux measured by Voyager 1 to $R^{\gamma}$ ($\gamma = -2$). The solid line (colored red in the electronic edition) is the formula in equation \ref{eqn:gtf}. A color version of this figure is available in the online journal.}
\label{fig:gtf}
\end{figure}

To describe the energy range spanned by the Voyager 1 data, we note that if we divide the Voyager 1 proton flux by a generic power law, as shown in Figure \ref{fig:gtf}, the resulting ratio looks like a sigmoid function in $\ln R$; we assume the following parametrization to describe this ratio:

\begin{equation} \label{eqn:gtf}
   \left[ 1 + \textrm{exp} \left( -\frac{\ln R - \mu}{\sigma} \right) \right]^{-1/\nu}
\end{equation}

where $\mu$ is related to the rigidity where the ratio is 1/2, $\sigma$ determines the steepness of the rise and $\nu$ describes a possible asymmetry.

The new parametrization for the LIS is therefore:

\begin{eqnarray} \label{eqn:lis}
   \begin{array}{l}
      \dfrac{dJ_{LIS}}{dR}(R) = N \left[ 1 + \textrm{exp} \left( -\dfrac{\ln R - \mu}{\sigma} \right) \right]^{-1/\nu} R^{\gamma_{1}} \\[2em]
      \times \left\{ 1 + \left[ \dfrac{R}{R_{b1}} \left( 1 + \left( \dfrac{R}{R_{b2}} \right)^{\Delta\gamma_{2}/s_{2}} \right)^{s_{2}}\right]^{\Delta\gamma_{1}/s_{1}} \right\}^{s_{1}}
   \end{array}
\end{eqnarray}

where $N$ is a normalization factor.

The available data in the rigidity range between up to a few tens of GV are all affected by the solar modulation, therefore a simple fit of equation \ref{eqn:lis} to Voyager 1 data and AMS-02 data above 100 GV (to remove any residual modulation) is not able to correctly constrain all the parameters, especially $\gamma_{1}$, $R_{b1}$, $\Delta\gamma_{1}$ and $s_{1}$. To resolve this issue, we proceed by simultaneously fitting Voyager 1 data with equation \ref{eqn:lis} and AMS-02 data with equation \ref{eqn:lis} modulated with the force-field approximation; this way, we obtain at the same time the parameters for the LIS and the average solar modulation parameter throughout the AMS-02 data time period. The least-squares fit is done with \texttt{MINUIT} \citep{bib:minuit}, minimizing the following quantity:

\begin{eqnarray} \label{eqn:glob-chisquare}
   \begin{array}{l}
      \chi^{2}_{glob} = \chi^{2}_{V1} + \chi^{2}_{AMS} \\
         = \displaystyle\sum_{i} \sigma^{-2}_{V1}(i) \left( y_{V1}(i) - \frac{1}{\Delta R_{i}} \!\!\! \displaystyle\int_{R_{i}}^{R_{i+1}} \!\!\! \frac{dJ_{LIS}}{dR}(R)dR \right)^{\!\!\!2} \\
         + \displaystyle\sum_{i} \sigma^{-2}_{AMS}(i) \left( y_{AMS}(i) - \frac{1}{\Delta R_{i}} \!\!\! \displaystyle\int_{R_{i}}^{R_{i+1}} \!\!\! \frac{dJ(R)}{dR}(R)dR \right)^{\!\!\!2}
   \end{array}
\end{eqnarray}

where $i$ is the binning index, $R_{i}$ and $R_{i+1}$ are the bin edges and $\Delta R_{i} = R_{i+1} - R_{i}$, $y(i)$ and $\sigma(i)$ are respectively the data and its associated error in the $i$-th bin and $dJ(R)/dR$ is defined as in equation \ref{eqn:ffa} after converting from kinetic energy to rigidity.

\begin{figure}[!h]
\centering
\includegraphics[width=0.475\textwidth]{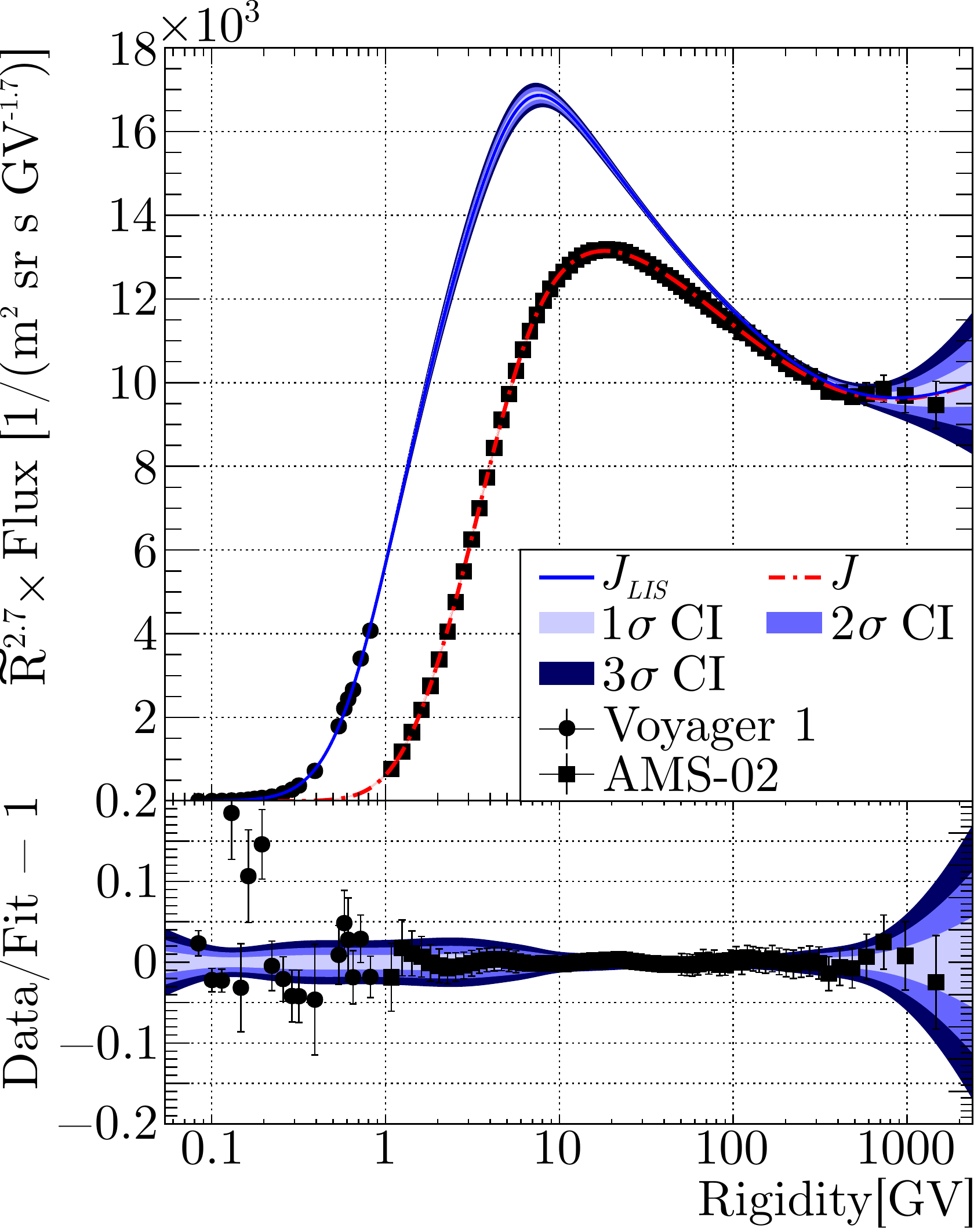}
\caption{The top panel shows the combined least-squares fit of Voyager 1 data (black dots) with $J_{LIS}$ (solid line) and AMS-02 data (black square) with $J$ (dashed-dotted line), as described in the text; the shaded bands represent the 1, 2 and 3 $\sigma$ confidence intervals around the best fit. Data and fit are rescaled by $\widetilde{R}^{2.7}$: see \citep{bib:stick-data-points} for the definition of $\widetilde{R}$. The bottom panel shows the fit residuals along with the confidence intervals. A color version of this figure is available in the online journal.}
\label{fig:lis-mod-fit-ffa}
\end{figure}

The results of the fit ($\chi^{2}_{glob}/\textrm{ndf} = 56/79$) are shown in Figure \ref{fig:lis-mod-fit-ffa}.

Voyager 1 errors are only statistical, so they over-constrain the LIS parameters; the residuals above $10\%$ in the bottom panel of Figure \ref{fig:lis-mod-fit-ffa} occur in the energy range where Voyager 1 changes its energy measurement method\footnote{Private communication with E. Stone and A. Cummings, 2015.}. The fitted parameters are presented in Table \ref{tab:ffa-fit-pars}.

\begin{table}[!h]
   \centering
   \begin{tabular}{lcc}
      \hline
      Parameter & Value & Error \\
      \hline
      \hline
      $N$ ($\textrm{m}^{-2} \textrm{ sr}^{-1} \textrm{ s}^{-1} \textrm{ GV}^{-1}$) & 11740 & $\pm 180$ \\
      $\mu$ & -0.559 & $\pm 0.011$ \\
      $\sigma$ & 0.563 & $\pm 0.005$ \\
      $\nu$ & 0.4315 & $\pm 0.0048$ \\
      $\gamma_{1}$ & -2.4482 & $\pm 0.0054$ \\
      $R_{b1}$ (GV) & 6.2 & $\pm 0.2$ \\
      $\Delta\gamma_{1}$ & -0.4227 & $\pm 0.0081$ \\
      $s_{1}$ & -0.108 & $\pm 0.015$ \\
      $R_{b2}$ (GV) & 545 & $\pm 210$ \\
      $\Delta\gamma_{2}$ & -0.6 & $\pm 0.2$ \\
      $s_{2}$ & -0.4 & $\pm 0.2$ \\
      \hline
      $\phi$ (MV) & 600 & $\pm 8$ \\
      \hline
   \end{tabular}
   \caption{Fitted parameters of the combined fit of Voyager 1 and AMS-02 data with the force-field approximation.}
   \label{tab:ffa-fit-pars}
\end{table}

\begin{figure}[!h]
\centering
\includegraphics[width=0.45\textwidth]{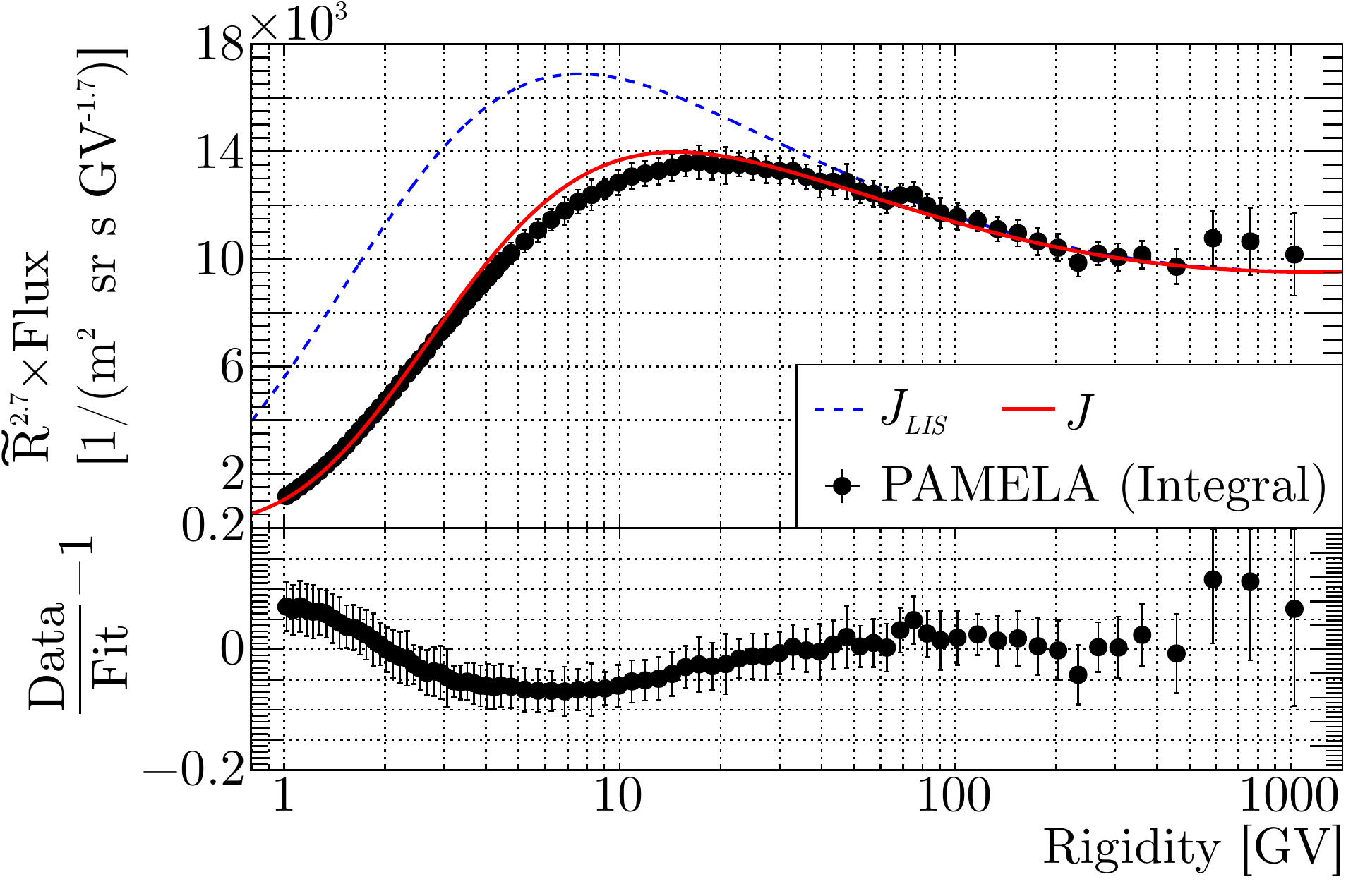}
\includegraphics[width=0.45\textwidth]{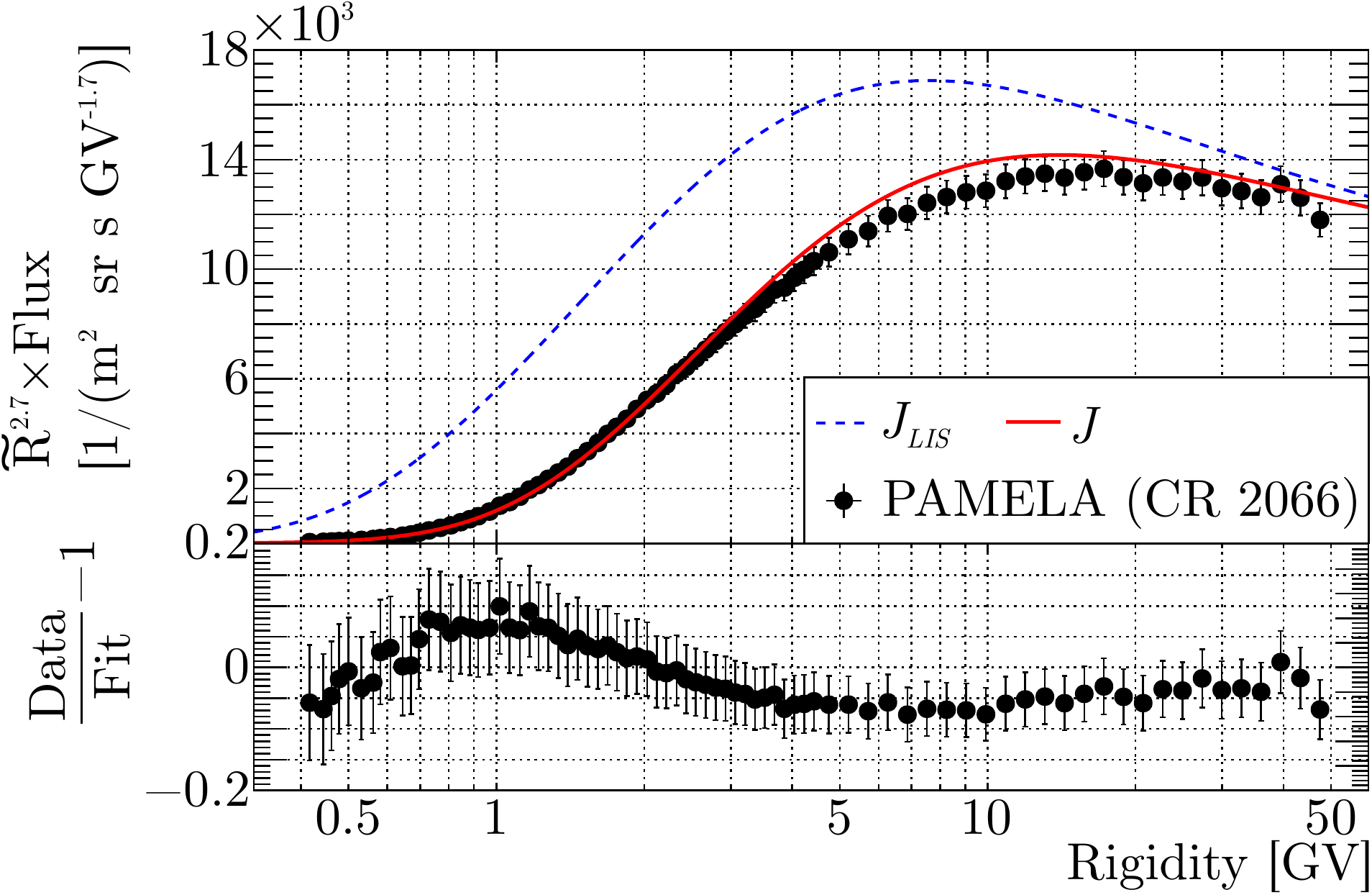}
\caption{Fit of the LIS modulated with the force-field approximation to the PAMELA integral proton flux (top figure) and to the PAMELA proton flux measured during the Carrington rotation 2066 (bottom figure). The dashed line represent the LIS, while the solid line is the modulated LIS fitted to the data (black dots). The lower panels in both figures show the fit residuals. A color version of this figure is available in the online journal.}
\label{fig:mod-fit-pamela-ffa}
\end{figure}

The PAMELA experiment published the proton flux between 0.4 GV and 50 GV, integrated in Carrington rotation periods, from July 2006 to January 2010 \footnote{Tables for all the Carrington rotation periods are available online in the COSMIC RAY database of the Italian Space Agency: http://tools.asdc.asi.it/cosmicRays.jsp} \citep{bib:pamela-monthly-flux}. This dataset provides valuable information for understanding the impact of the solar modulation on the differential flux.
Using the force-field approximation from equation \ref{eqn:ffa}, we fit the PAMELA data with the LIS in equation \ref{eqn:lis}. In order to take into account the uncertainty on the LIS in the error on the fitted $\phi$, we also fit the PAMELA data with the LIS plus or minus the $1 \sigma$ confidence interval, thus getting $\phi_{\pm1}$, and we take the difference $\phi - \phi_{\pm1}$ as an estimate of the LIS uncertainty propagated to the fitted modulation parameter.
Figure \ref{fig:mod-fit-pamela-ffa} illustrates an example of the fit results for the proton flux measured by PAMELA between July 2006 and March 2008 (top) and during Carrington rotation 2066 (bottom); Table \ref{tab:phi-ffa-dff}, columns 2 to 4, presents the fitted values of $\phi$ with the errors coming from the fit itself and from the LIS.

Although the reduced chi-square is good for both fits (respectively $79/79$ and $55/77$), the residuals have a structure with a bump around 1 GV and a dip around 7 GV, meaning that the fit does not completely describe the data. The same behavior is observed in the residuals of the fits to all PAMELA monthly fluxes, with the bump and the dip occurring around the same rigidities. We believe that these structures are due to the fact that the force-field approximation does not correctly reproduce the solar modulation during the minimum of solar cycle 23 because some processes (like drift) are not considered. \citep{bib:drift}.\vspace*{0.3cm}

\section{Beyond the force-field approximation} \label{sec:beyond-ffa}
These results suggest that the solar modulation may affect GCRs below and above a few GVs in different ways; a similar conclusion is also found in \citep{bib:gieseler} by comparing data from NMs, PAMELA and the EPHIN instrument on board the SOHO spacecraft.
To account for this effect, we modify the force-field approximation by considering an energy-dependent solar modulation parameter:

\begin{eqnarray} \label{eqn:dff}
   \phi(T) & = & \left\{
   \begin{array}{ll}
      \phi_{L}, & T < T_{L} \\
      f(T,\,\phi_{L},\,\phi_{H}), & T_{L} \leq T \leq T_{H} \\
      \phi_{H}, & T > T_{H}
   \end{array}
   \right.
\end{eqnarray}

where the indices $L$ and $H$ stand for ``low'' and ``high'' energy and $f$ is a transition function between $\phi_{L}$ and $\phi_{H}$. We want $f$ to have a zero derivative at $T_{L}$ and $T_{H}$ to avoid discontinuities in the spectral index: the simplest function that has this property is a third degree polynomial, which is completely constrained by the given boundary conditions. Defining $t = (T - T_{L})/(T_{H} - T_{L})$, the transition function is $f(T,\,\phi_{L},\,\phi_{H}) = \phi_{L} + (\phi_{H} - \phi_{L})t^{2}(3-2t)$.


We then proceed as previously done: simultaneously fitting Voyager 1 and AMS-02 data, minimizing the global chi-square defined in equation \ref{eqn:glob-chisquare} and replacing the solar modulation parameter in equation \ref{eqn:ffa} with the one defined in equation \ref{eqn:dff}.

\begin{figure}[!h]
\centering
\includegraphics[width=0.475\textwidth]{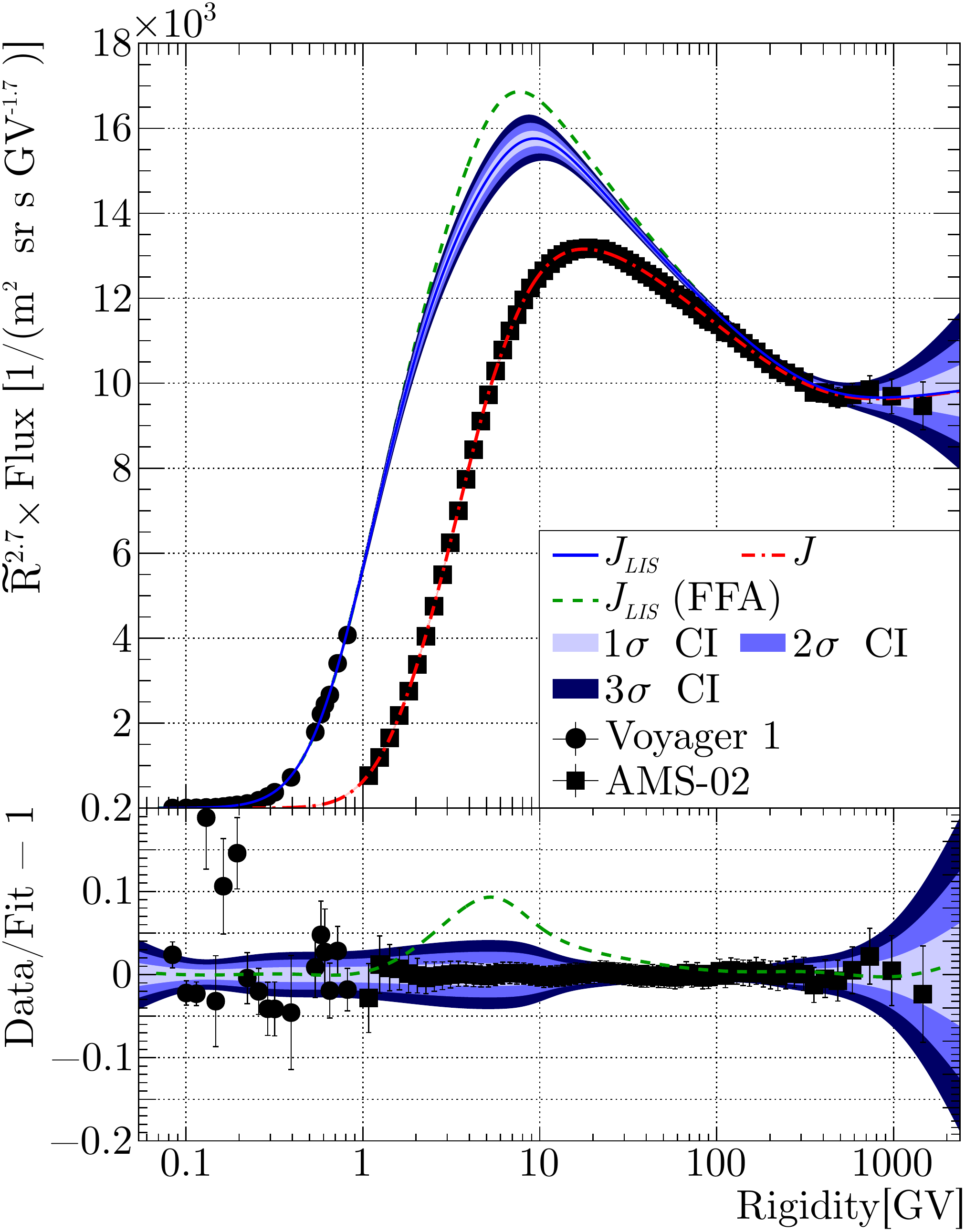}
\caption{Same as Figure \ref{fig:lis-mod-fit-ffa}, but using the energy dependent solar modulation parameter of equation \ref{eqn:dff}. For comparison, the LIS derived in Section \ref{sec:new-lis} with the usual force-field approximation, $J_{LIS}$ (FFA), is shown as a dashed line. A color version of this figure is available in the online journal.}
\label{fig:lis-mod-fit-dff}
\end{figure}

The results of the fit ($\chi^{2}_{glob}/\textrm{ndf} = 55/78$) are shown in Figure \ref{fig:lis-mod-fit-dff}.

In the fit, the values of $T_{L}$ and $T_{H}$ have been fixed at 0.125 and 4.65 GeV, while the fitted parameters are presented in Table \ref{tab:dff-fit-pars}. The asymptotic spectral indices at intermediate and high rigidities are, respectively, $\gamma_{2} = -2.853 \pm 0.015$ and $\gamma_{3} = -2.674 \pm 0.073$.

\begin{table}[!h]
   \centering
   \begin{tabular}{lcc}
      \hline
      Parameter & Value & Error \\
      \hline
      \hline
      $N$ ($\textrm{m}^{-2} \textrm{ sr}^{-1} \textrm{ s}^{-1} \textrm{ GV}^{-1}$) & 13020 & $\pm 240$ \\
      $\mu$ & -0.526 & $\pm 0.011$ \\
      $\sigma$ & 0.579 & $\pm 0.005$ \\
      $\nu$ & 0.4052 & $\pm 0.0046$ \\
      $\gamma_{1}$ & -2.5794 & $\pm 0.0059$ \\
      $R_{b1}$ (GV) & 8.69 & $\pm 0.49$ \\
      $\Delta\gamma_{1}$ & -0.2735 & $\pm 0.0089$ \\
      $s_{1}$ & -0.068 & $\pm 0.016$ \\
      $R_{b2}$ (GV) & 410 & $\pm 190$ \\
      $\Delta\gamma_{2}$ & -0.65 & $\pm 0.29$ \\
      $s_{2}$ & -0.27 & $\pm 0.24$ \\
      \hline
      $\phi_{L}$ (MV) & 589 & $\pm 8$ \\
      $\phi_{H}$ (MV) & 485 & $\pm 22$ \\
      \hline
   \end{tabular}
   \caption{Fitted parameters of the combined fit of Voyager 1 and AMS-02 data with the modified force-field approximation.}
   \label{tab:dff-fit-pars}
\end{table}

\begin{figure}[!h]
\centering
\includegraphics[width=0.45\textwidth]{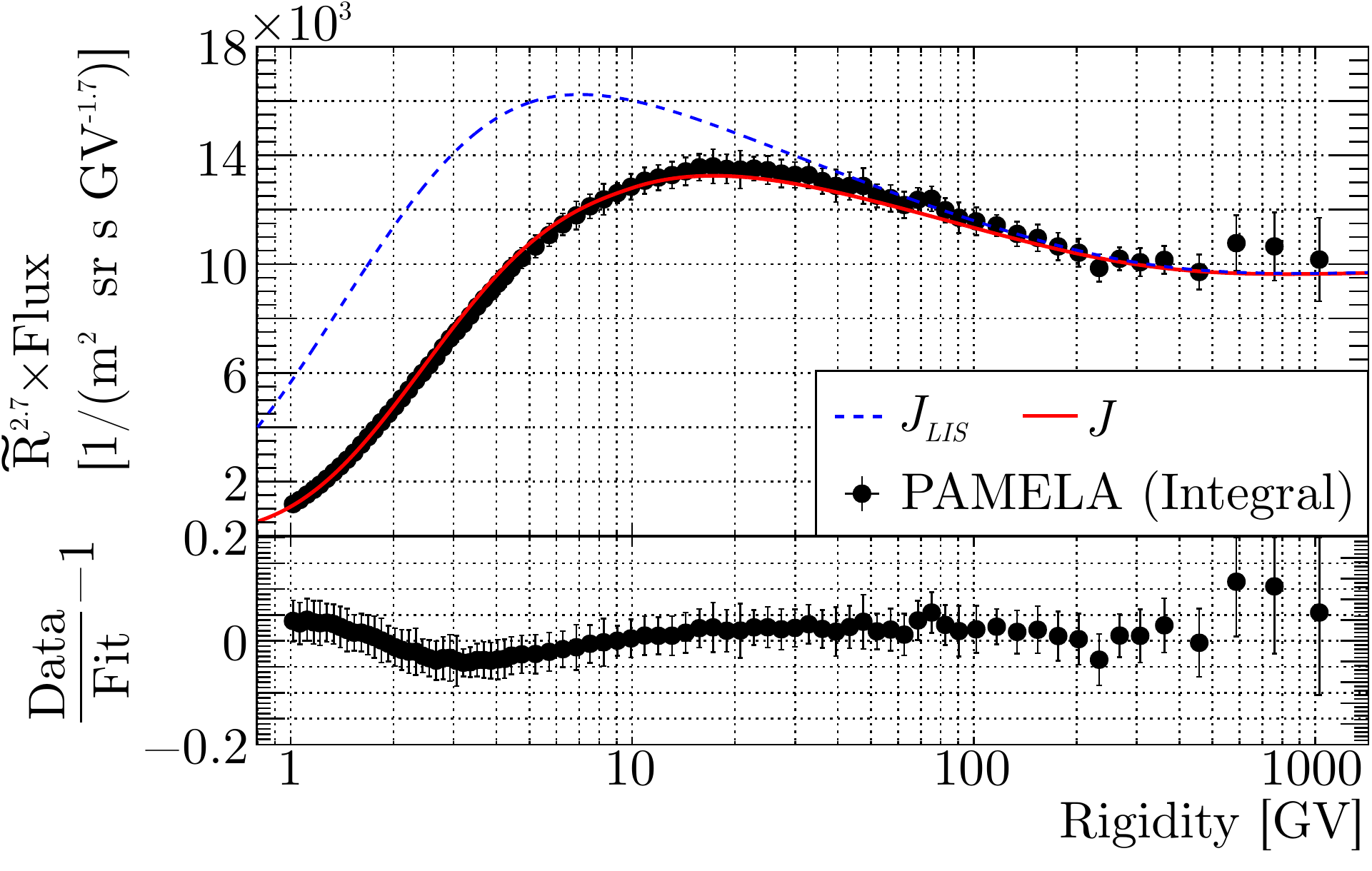}
\includegraphics[width=0.45\textwidth]{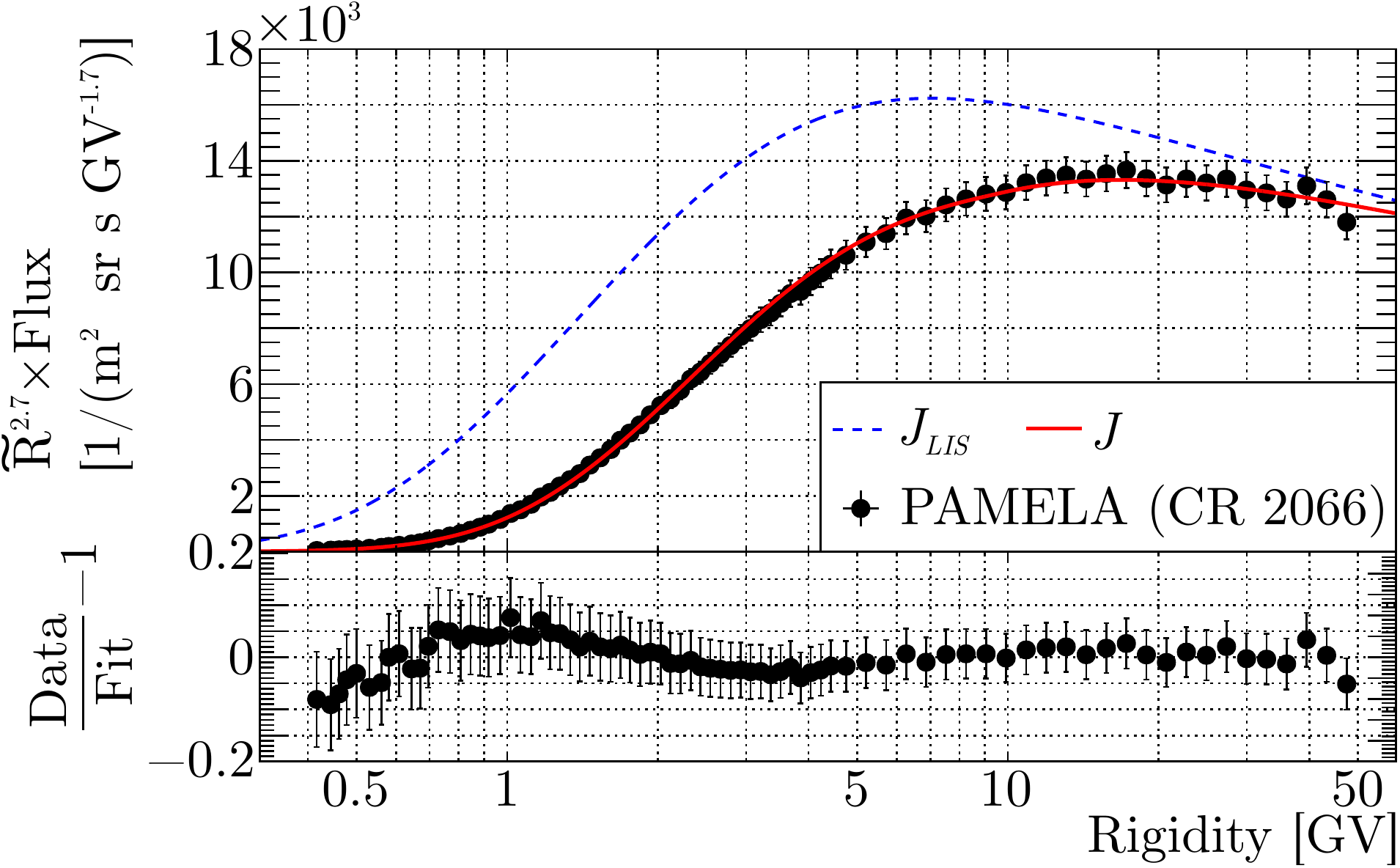}
\caption{Same as Figure \ref{fig:mod-fit-pamela-ffa}, but using the energy dependent solar modulation parameter of equation \ref{eqn:dff}. A color version of this figure is available in the online journal.}
\label{fig:mod-fit-pamela-dff}
\end{figure}

Figure \ref{fig:mod-fit-pamela-dff} shows the proton flux measured by PAMELA during the same time periods shown in Figure \ref{fig:mod-fit-pamela-ffa} fitted with the energy dependent solar modulation parameter: the reduced chi-squares are, respectively, $13/78$ and $36/76$ and the structures of the residuals are now smaller with respect to the ones obtained with the force-field approximation.

\begin{figure}[!b]
\centering
\includegraphics[width=0.45\textwidth]{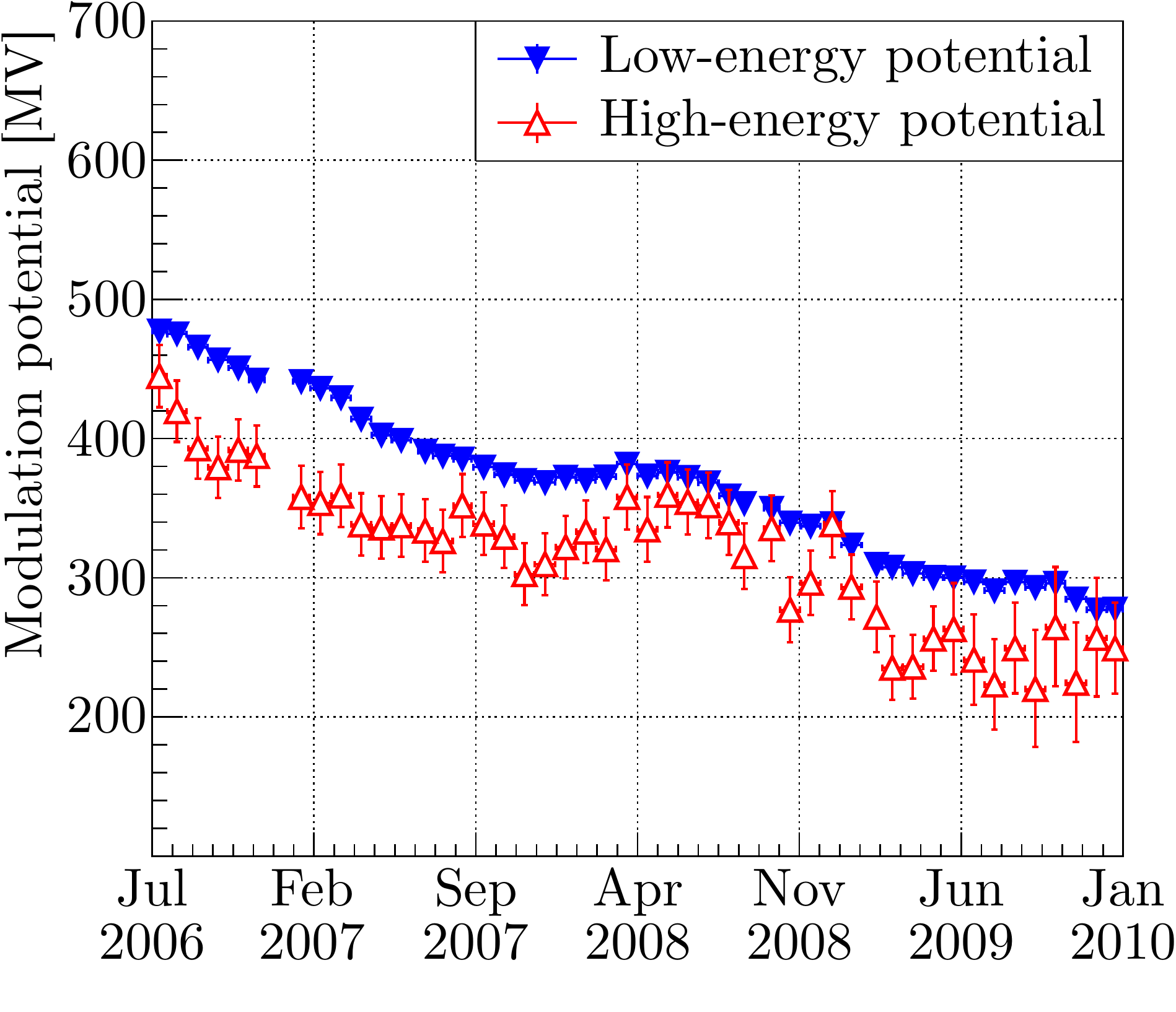}
\includegraphics[width=0.45\textwidth]{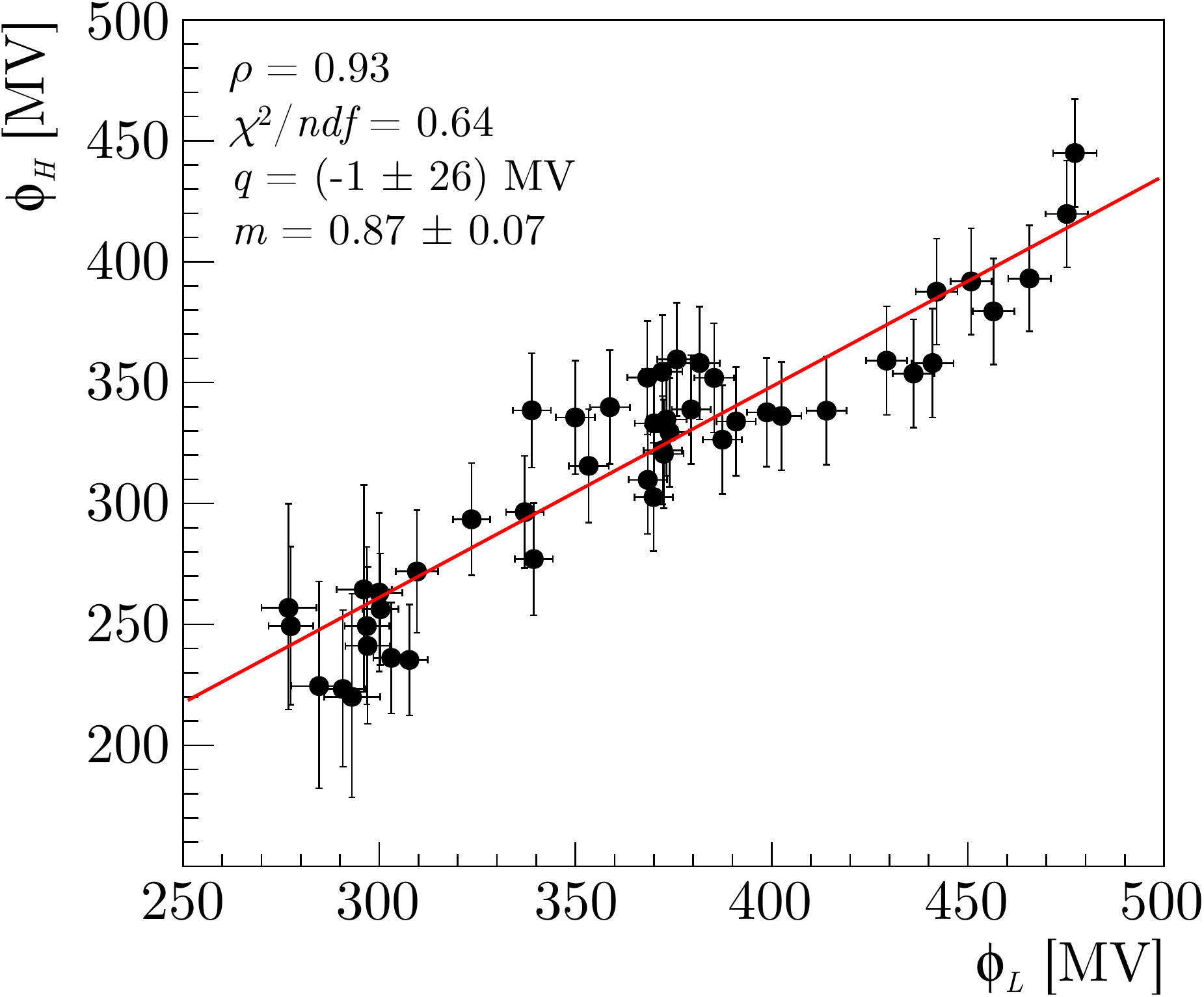}
\includegraphics[width=0.45\textwidth]{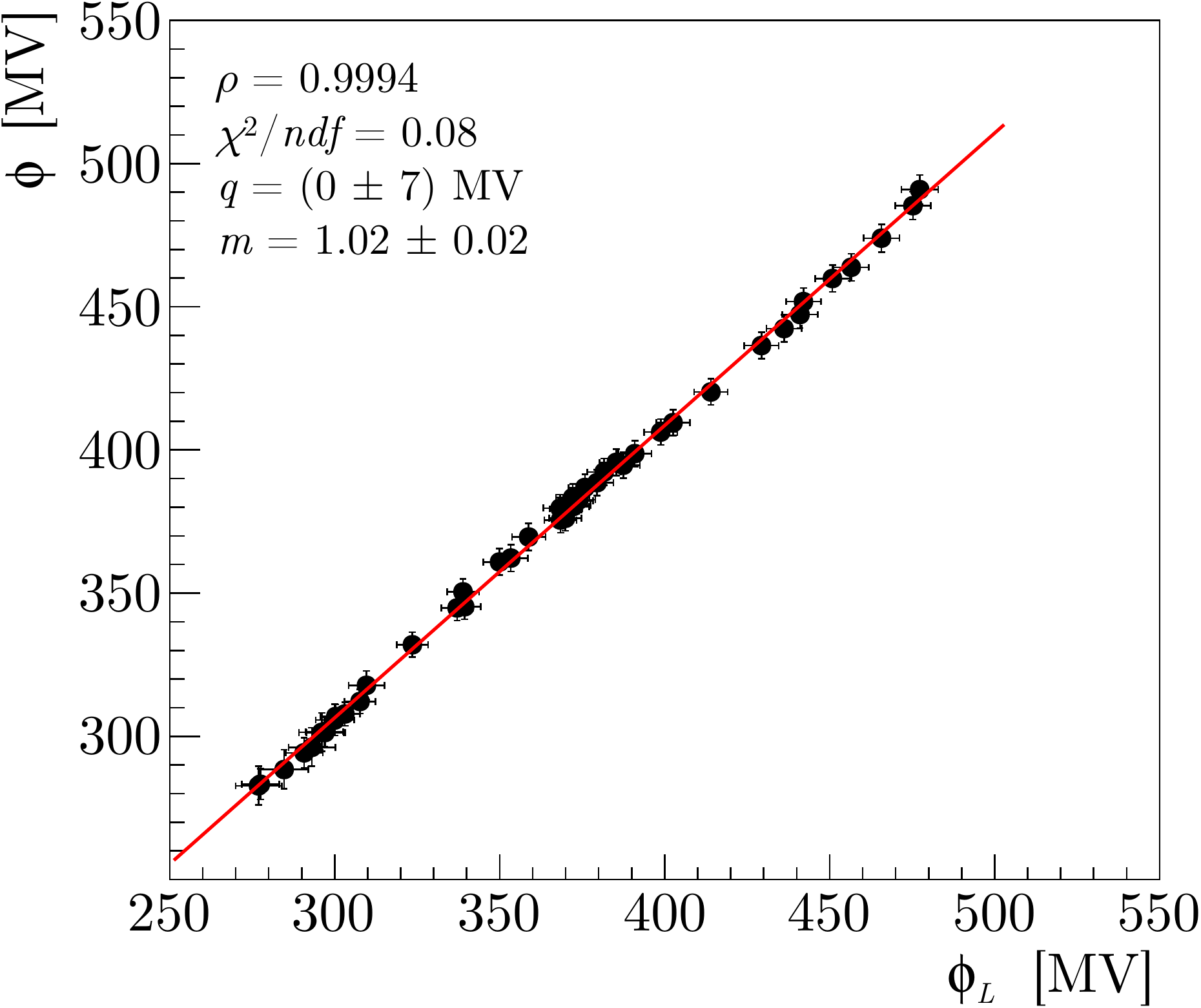}
\caption{(Top) Time dependence of the solar modulation parameters $\phi_{L}$ (down triangles) and $\phi_{H}$ (up triangles) derived from the monthly proton fluxes measured by PAMELA during the minimum of solar cycle 23. (Center) Correlation between $\phi_{L}$ and $\phi_{H}$; the solid line is a linear fit. (Bottom) Correlation between $\phi_{L}$ and $\phi$, the solar modulation parameter obtained with the force-field approximation. A color version of this figure is available in the online journal.}
\label{fig:phi-dff-pamela}
\end{figure}

The fit has been repeated for all monthly proton fluxes measured by PAMELA and the time dependence of the fitted solar modulation parameters $\phi_{L}$ and $\phi_{H}$ is plotted in the top panel of Figure \ref{fig:phi-dff-pamela}. The fitted values of $\phi_{L}$ and $\phi_{H}$ are presented in Table \ref{tab:phi-ffa-dff}, columns 5 to 10. The two modulation parameters are well-correlated (the correlation coefficient is $\rho = 0.93$), as shown in the central panel of Figure \ref{fig:phi-dff-pamela}. A linear fit to the modulation parameters has been performed, yielding a $\chi^{2}/ndf = 0.64$ with a slope of $0.87 \pm 0.07$ and an intercept compatible with zero. If we interpret the solar modulation parameters as the average energy losses experienced by the particles traveling from the edge of the heliosphere up to the Earth, these results show that, during the minimum of solar cycle 23, the energy losses are slightly higher at lower rigidities, while the force-field approximation predicts the same energy loss at all rigidities. The bottom panel of Figure \ref{fig:phi-dff-pamela} shows the correlation between $\phi_{L}$ and the $\phi$ previously obtained with the force-field approximation: the correlation coefficient is 0.9994 and a linear fit yields a $\chi^{2}/ndf = 0.08$ with a slope of $1.02 \pm 0.02$ and an intercept compatible with zero. This result means that the force-field approximation is able to capture the leading effects of the solar modulation down to 0.5 GV even when the assumptions of the approximation are not completely satisfied, such as during a solar minimum.

\section{Comparison with neutron monitors} \label{sec:nm-comparison}
The effect of the solar modulation on GCRs has been continuously measured on ground since the 1950's with the world network of NMs, which measure the integral of the GCR flux above the rigidity cutoff pertaining to the NM location. In order to extract the solar modulation parameter from NM data, the shape of the LIS and the elemental composition of GCRs must be assumed, usually from measurements made by balloon- and space-borne experiments.

\begin{figure}[!h]
\centering
\includegraphics[width=0.45\textwidth]{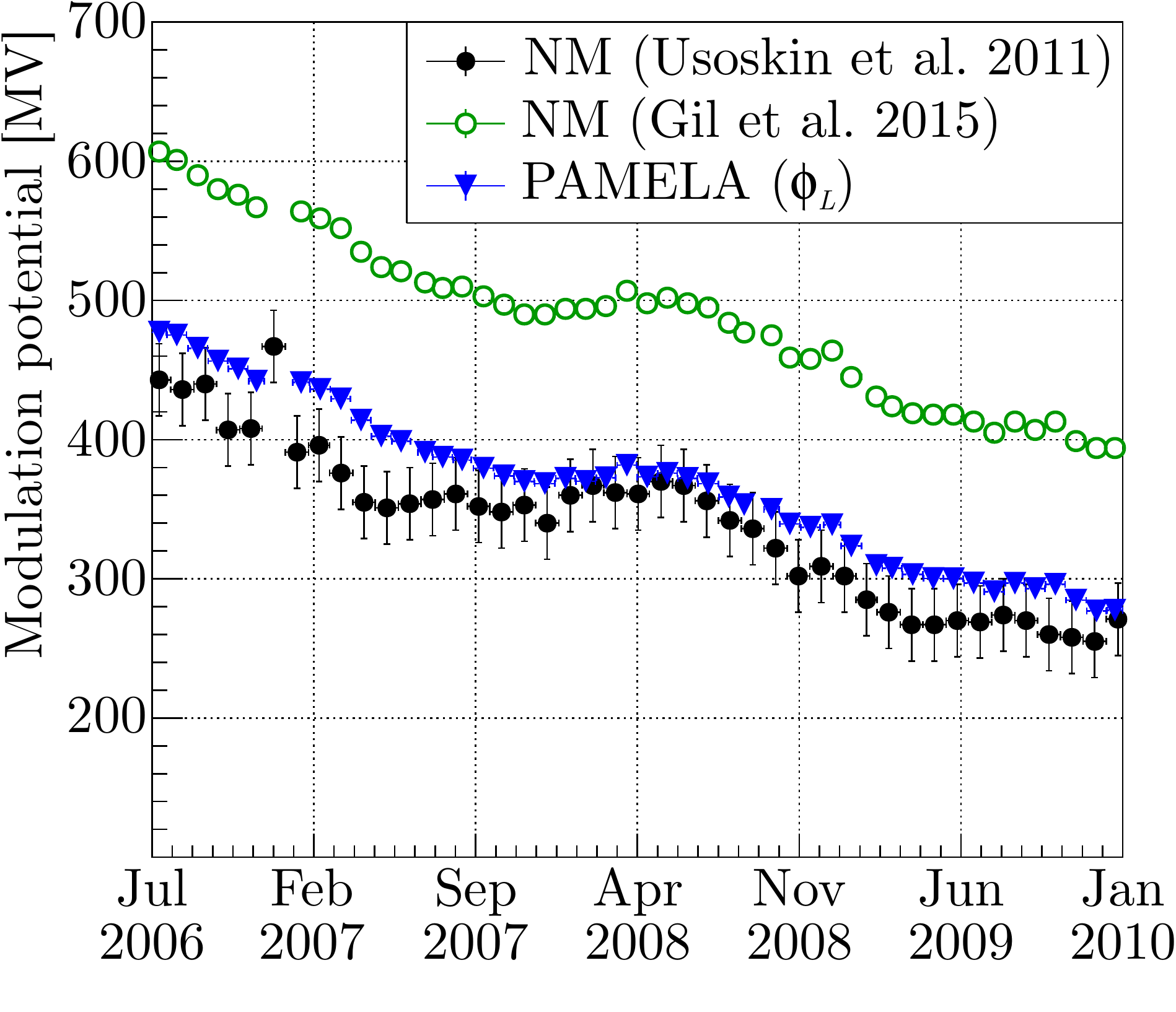}
\includegraphics[width=0.45\textwidth]{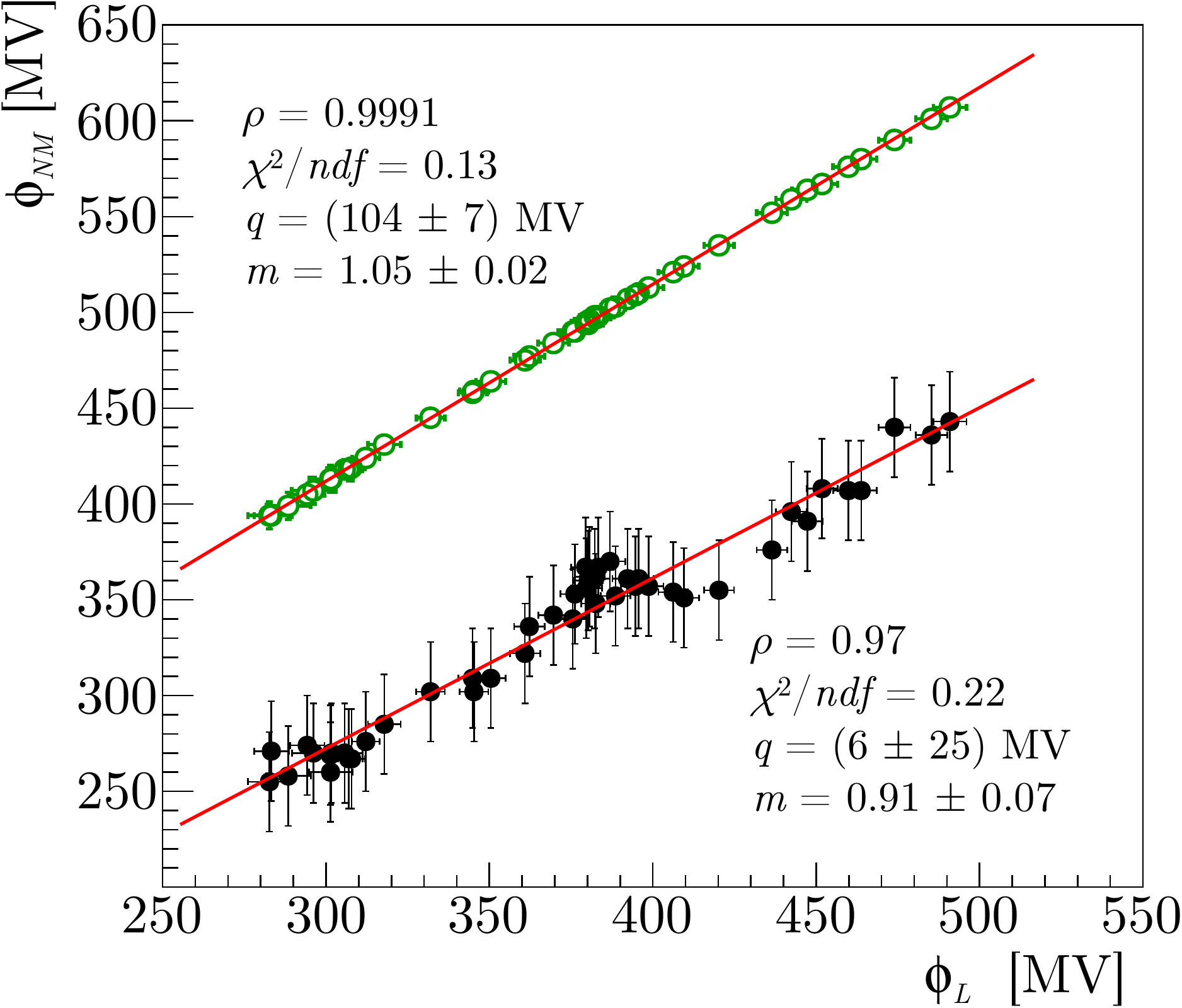}
\caption{(Top) Low energy solar modulation parameter extracted from PAMELA monthly proton fluxes (down triangles) compared with the one derived from neutron monitors (filled and hollow dots). (Bottom) Correlation between $\phi_{L}$ and $\phi_{NM}$; the solid lines are linear fits. A color version of this figure is available in the online journal.}
\label{fig:phi-pamela-nm}
\end{figure}

Figure \ref{fig:phi-pamela-nm} top shows the comparison between the low-energy solar modulation parameter, $\phi_{L}$, obtained from the fits of equation \ref{eqn:lis} to the PAMELA proton fluxes and the parameter derived from NMs, $\phi_{NM}$ \citep{bib:Usoskin11,bib:gil}, using the BPH00 model as the LIS (see Figure \ref{fig:exp-lis-ratio}).

With respect to \citet{bib:Usoskin11}, \citet{bib:gil} use a new improved yield function that takes into account the effect of the finite lateral size of the atmospheric showers induced by GCRs \citep{bib:mishev} and calibrate the normalization of different NM stations with the proton monthly fluxes measured by PAMELA. The very small difference between $\phi_{L}$ and $\phi_{NM}$ from \citep{bib:Usoskin11} is by accident, since in this case, in addition to the different LIS used to fit the data, the yield function did not include the effect described in \citep{bib:mishev}. The comparison with $\phi_{NM}$ from \citep{bib:gil} results in a better correlation ($\rho = 0.9991$ versus $\rho = 0.97$) and the shift of $\approx 100$ MV is due only to the different LIS adopted in \citep{bib:gil} with respect to the one used in this work. Figure \ref{fig:phi-pamela-nm} bottom shows the correlation between $\phi_{L}$ and $\phi_{NM}$ from the two cited works. Assuming that the linear relation between $\phi_{L}$ and $\phi_{NM}$, shown in Figure \ref{fig:phi-pamela-nm} bottom, holds also for different periods of solar modulation, we can use the following expressions to compute the modulation parameter associated with the new parametrization of the proton LIS throughout the whole period of data taking of NMs:

\begin{eqnarray}
   \begin{array}{l}
      \phi_{L} = (1.10 \pm 0.08) \phi_{N\!M,Usoskin} - (7 \pm 27) \textrm{ MV} \\
      \phi_{L} = (0.95 \pm 0.08) \phi_{N\!M,Gil} - (99 \pm 7) \textrm{ MV}.
   \end{array}
\end{eqnarray}

\section{Conclusions}
A new parametrization of the proton LIS from 80 MV up to 2 TV has been derived in this work using the measurements of the proton flux performed by Voyager 1 and AMS-02. The LIS is characterized by two power-laws with breaks at $(8.69 \pm 0.49)$ GV and $(410 \pm 190)$ GV with a spectral index changing from $-2.5794 \pm 0.0059$ to $-2.853 \pm 0.015$ at low energy and to $-2.674 \pm 0.073$ at high energy. The force-field approximation is not able to accurately describe the solar modulation measured by PAMELA during the minimum of solar cycle 23, therefore we introduce an energy-dependent modulation parameter that yields a better result for the PAMELA data. A linear relation between the published values of the modulation parameter derived from NMs and the one obtained in this work is given. With the availability of precise measurements directly from space, we can finally start to understand the details of the processes that cause the solar modulation of galactic cosmic rays. It would be interesting to see how the monthly proton fluxes measured by AMS-02 during the current solar maximum compares with the results presented in this work.\\

{\sc Note added.} During the completion of this work, we became aware of a related study by Ghelfi, Barao, Derome and Maurin. It focuses on the determination of interstellar proton and helium flux with splines. Both the data sets and methods used differ from those of our study, making the two analyses complementary. A comparison of their proton LIS (obtained with the force-field approximation) and ours (obtained with the energy-dependent solar modulation parameter) shows a very good agreement in the range 4~GV~--~1~TeV, but with different uncertainties.

\acknowledgments
We would like to thank E. Stone and A. Cummings for the discussion about the Voyager 1 data; M. Potgieter for the fruitful discussion about the theoretical interpretation; I. Usoskin for the help with the neutron monitor comparison.

This work has been founded by: National Science Foundation Early Career under grant (NSF 1455202); Wyle Laboratories, Inc.; NASA and Earth Space Science Fellowship under grant (15-HELIO15F-0005); Research Corporation University of Hawaii.

\appendix

\section{Solar modulation parameters from PAMELA protons fluxes}

\begin{table}[!h]
   \centering
   \begin{tabular}{c|ccc|ccc|ccc}
      \hline
      Date & $\phi$ & $\sigma_{fit}$ & $\sigma_{LIS}$ & $\phi_{L}$ & $\sigma_{fit}$ & $\sigma_{LIS}$ & $\phi_{H}$ & $\sigma_{fit}$ & $\sigma_{LIS}$\\
      \hline
      \hline
      2006/07/07 -- 2006/07/26 & 490.9 & 3.8 & 3.4 & 477.3 & 4.0 & 4.0 & 445 & 14 & 17\\
      2006/07/27 -- 2006/08/22 & 485.3 & 3.6 & 3.3 & 475.2 & 3.7 & 3.9 & 420 & 14 & 17\\
      2006/08/24 -- 2006/09/19 & 474.0 & 3.6 & 3.3 & 465.7 & 3.8 & 3.9 & 393 & 14 & 17\\
      2006/09/20 -- 2006/10/16 & 463.8 & 3.5 & 3.2 & 456.5 & 3.7 & 3.8 & 379 & 14 & 17\\
      2006/10/17 -- 2006/11/12 & 459.9 & 3.5 & 3.2 & 450.8 & 3.7 & 3.8 & 392 & 14 & 17\\
      2006/11/13 -- 2006/12/04 & 451.8 & 3.5 & 3.2 & 442.1 & 3.7 & 3.8 & 388 & 14 & 17\\
      2007/01/11 -- 2007/02/02 & 447.3 & 3.6 & 3.1 & 441.0 & 3.8 & 3.8 & 358 & 15 & 17\\
      2007/02/03 -- 2007/03/02 & 442.4 & 3.6 & 3.1 & 436.2 & 3.8 & 3.7 & 354 & 15 & 17\\
      2007/03/03 -- 2007/03/29 & 436.5 & 3.5 & 3.1 & 429.3 & 3.7 & 3.7 & 359 & 15 & 17\\
      2007/03/30 -- 2007/04/25 & 420.2 & 3.5 & 3.0 & 414.0 & 3.6 & 3.6 & 338 & 15 & 17\\
      2007/04/26 -- 2007/05/22 & 409.5 & 3.5 & 3.0 & 402.5 & 3.6 & 3.6 & 336 & 15 & 17\\
      2007/05/23 -- 2007/06/17 & 406.2 & 3.4 & 3.0 & 398.8 & 3.6 & 3.6 & 338 & 15 & 17\\
      2007/06/27 -- 2007/07/16 & 398.7 & 3.5 & 3.0 & 390.9 & 3.6 & 3.5 & 334 & 15 & 17\\
      2007/07/17 -- 2007/08/12 & 394.7 & 3.4 & 2.9 & 387.5 & 3.6 & 3.5 & 326 & 15 & 17\\
      2007/08/13 -- 2007/09/06 & 395.7 & 3.5 & 3.0 & 385.3 & 3.7 & 3.5 & 352 & 15 & 17\\
      2007/09/09 -- 2007/10/06 & 388.6 & 3.4 & 3.0 & 379.5 & 3.5 & 3.5 & 339 & 15 & 17\\
      2007/10/07 -- 2007/11/02 & 382.6 & 3.4 & 2.9 & 374.0 & 3.5 & 3.5 & 329 & 15 & 17\\
      2007/11/03 -- 2007/11/29 & 376.1 & 3.3 & 2.9 & 369.9 & 3.5 & 3.4 & 303 & 15 & 17\\
      2007/11/30 -- 2007/12/27 & 375.5 & 3.4 & 2.9 & 368.4 & 3.5 & 3.4 & 310 & 15 & 17\\
      2007/12/28 -- 2008/01/23 & 380.2 & 3.4 & 2.9 & 372.2 & 3.5 & 3.5 & 322 & 15 & 17\\
      2008/01/24 -- 2008/02/19 & 379.4 & 3.4 & 3.0 & 370.1 & 3.5 & 3.5 & 333 & 15 & 17\\
      2008/02/20 -- 2008/03/17 & 380.6 & 3.5 & 3.0 & 372.6 & 3.6 & 3.5 & 321 & 15 & 17\\
      2008/03/19 -- 2008/04/14 & 392.3 & 3.6 & 3.0 & 381.7 & 3.7 & 3.5 & 358 & 16 & 17\\
      2008/04/15 -- 2008/05/11 & 382.2 & 3.6 & 2.9 & 373.2 & 3.7 & 3.5 & 335 & 16 & 17\\
      2008/05/12 -- 2008/06/07 & 386.9 & 3.6 & 3.0 & 375.9 & 3.7 & 3.5 & 360 & 16 & 17\\
      2008/06/08 -- 2008/07/04 & 383.3 & 3.7 & 3.0 & 372.1 & 3.8 & 3.5 & 354 & 16 & 17\\
      2008/07/05 -- 2008/08/01 & 379.7 & 3.7 & 3.0 & 368.3 & 3.8 & 3.5 & 352 & 16 & 17\\
      2008/08/02 -- 2008/08/28 & 369.6 & 3.7 & 2.9 & 358.8 & 3.8 & 3.4 & 340 & 16 & 17\\
      2008/08/29 -- 2008/09/11 & 362.2 & 3.7 & 2.9 & 353.4 & 3.8 & 3.4 & 316 & 16 & 17\\
      2008/10/01 -- 2008/10/21 & 360.9 & 3.6 & 2.9 & 350.0 & 3.7 & 3.4 & 336 & 16 & 17\\
      2008/10/22 -- 2008/11/18 & 345.3 & 3.4 & 2.7 & 339.4 & 3.6 & 3.3 & 277 & 16 & 17\\
      2008/11/19 -- 2008/12/15 & 344.8 & 3.4 & 2.8 & 337.1 & 3.5 & 3.3 & 296 & 16 & 17\\
      2008/12/20 -- 2009/01/11 & 350.4 & 3.6 & 2.9 & 338.9 & 3.6 & 3.3 & 338 & 16 & 17\\
      2009/01/12 -- 2009/02/08 & 332.0 & 3.4 & 2.8 & 323.5 & 3.5 & 3.2 & 293 & 16 & 17\\
      2009/02/21 -- 2009/03/07 & 317.8 & 4.2 & 2.8 & 309.6 & 4.3 & 3.3 & 272 & 19 & 17\\
      2009/03/08 -- 2009/04/03 & 312.2 & 3.3 & 2.6 & 307.7 & 3.5 & 3.2 & 235 & 16 & 17\\
      2009/04/04 -- 2009/05/01 & 307.8 & 3.2 & 2.6 & 303.1 & 3.4 & 3.1 & 236 & 16 & 17\\
      2009/05/02 -- 2009/05/28 & 306.9 & 3.3 & 2.6 & 300.3 & 3.4 & 3.1 & 256 & 16 & 17\\
      2009/05/29 -- 2009/06/24 & 305.7 & 4.9 & 2.5 & 300.0 & 5.0 & 3.1 & 263 & 28 & 17\\
      2009/06/25 -- 2009/07/21 & 301.3 & 4.7 & 2.4 & 297.0 & 4.8 & 3.0 & 241 & 28 & 17\\
      2009/07/22 -- 2009/08/18 & 294.2 & 4.7 & 2.4 & 290.7 & 4.8 & 3.0 & 223 & 28 & 17\\
      2009/08/19 -- 2009/09/14 & 301.6 & 4.8 & 2.4 & 296.9 & 4.9 & 3.0 & 249 & 28 & 17\\
      2009/09/15 -- 2009/10/11 & 296.1 & 6.3 & 2.4 & 293.1 & 6.5 & 3.0 & 220 & 39 & 17\\
      2009/10/12 -- 2009/11/07 & 301.4 & 6.3 & 2.4 & 296.1 & 6.4 & 3.0 & 264 & 39 & 17\\
      2009/11/08 -- 2009/12/05 & 288.4 & 6.4 & 2.3 & 284.7 & 6.6 & 3.0 & 224 & 40 & 17\\
      2009/12/06 -- 2010/01/01 & 282.7 & 6.3 & 2.4 & 276.9 & 6.4 & 2.9 & 257 & 39 & 17\\
      2010/01/02 -- 2010/01/23 & 283.3 & 4.8 & 2.4 & 277.5 & 4.9 & 3.0 & 249 & 28 & 17\\
      \hline
   \end{tabular}
   \caption{Solar modulation parameter in units of MV derived from PAMELA monthly proton fluxes. The error contributions from the fit of the PAMELA fluxes ($\sigma_{fit}$) and from the uncertainty on the LIS ($\sigma_{LIS}$) are reported separately. $\phi$ is the modulation parameter obtained with the force-field approximation, while $\phi_{L}$ and $\phi_{H}$ are the modulation parameters obtained with the modified force-field approximation.}
   \label{tab:phi-ffa-dff}
\end{table}

\end{document}